\documentclass[11pt]{article}
\pdfoutput=1

\usepackage{bm}
\usepackage[normalem]{ulem}
\usepackage[utf8]{inputenc}
\usepackage[T1]{fontenc}
\usepackage{mathptmx}
\usepackage{etoolbox}
\usepackage{xcolor}
\usepackage{physics}

\usepackage[a4paper, margin=0.7in]{geometry}
\usepackage{authblk}
\usepackage{graphicx}
\usepackage{dcolumn}
\usepackage{amsmath}
\usepackage{amssymb}
\usepackage{textcomp}
\usepackage{verbatim}
\usepackage{float}
\usepackage[symbol]{footmisc}
\setfnsymbol{wiley}

\usepackage{multirow}
\usepackage[numbers,sort&compress]{natbib}

\usepackage{hyperref}
\hypersetup{
    colorlinks=true,
    linkcolor=blue,
    filecolor=blue,
    urlcolor=blue,
    citecolor = blue,
}

\begin{document}
\title{Design of Cost-Effective Nanoacoustic Devices based on Mesoporous Thin Films}
\author[1]{E. R. Cardozo de Oliveira}
\author[2]{P. Vensaus}
\author[2]{G. J. A. A. Soler-Illia}
\author[1]{N. D. Lanzillotti-Kimura}

\affil[1]{Université Paris-Saclay, CNRS, Centre de Nanosciences et de Nanotechnologies, 91120 Palaiseau, France}

\affil[2]{Instituto de Nanosistemas, Escuela de Bio y Nanotecnologías, Universidad Nacional de San Martín-CONICET, Buenos Aires, Argentina}

\date{}
\maketitle 
\begin{abstract}
Gigahertz acoustic resonators have the potential to advance data processing and quantum communication. However, they are expensive and lack responsiveness to external stimuli, limiting their use in sensing applications. In contrast, low-cost nanoscale mesoporous materials, known for their high surface-to-volume ratio, have shown promise in various applications. We recently demonstrated that mesoporous silicon dioxide (SiO$_{2}$) and titanium dioxide (TiO$_{2}$) thin layers can support coherent acoustic modes in the 5 to 100 GHz range. In this study, we propose a new method for designing tunable acoustic resonators using mesoporous thin films on acoustic distributed Bragg reflectors.  By infiltrating the pores with different chemicals, the material's properties could be altered and achieve tunability in the acoustic resonances. We present four device designs and use simulations to predict resonators with Q-factors up to 1000. We also observe that the resonant frequency and intensity show a linear response to relative humidity, with a tunability of up to 60$\%$. Our platform offers a unique  opportunity to design cost-effective nanoacoustic sensing and reconfigurable optoacoustic nanodevices.

\end{abstract}

\section{Introduction}

The potential of nanophononics -- the engineering and manipulation of acoustic phonons at the nanoscale --,~\cite{priya_perspectives_2023,balandin_nanophononics_2005,Volz2016,trigo_confinement_2002,anguiano_micropillar_2017,lanzillotti-kimura_nanowave_2006-1,lanzillotti-kimura_acoustic_2007} has been established within various technological applications, including heat management, data processing, and quantum technologies. The field explores how the interaction between GHz acoustic phonons and other excitations, dependent on the position of the atoms in a lattice, can be leveraged to engineer novel and more efficient devices in phonon-based optoelectronics,~\cite{stiller_coherently_2020,dainese_stimulated_2006,de_lima_compact_2006,young_subterahertz_2012,dunn_high-speed_2020} photonics,~\cite{trigo_confinement_2002,Fainstein2007,fainstein_strong_2013,lanzillotti-kimura_enhanced_2011,ortiz_fiber-integrated_2020,ortiz_topological_2021-1,czerniuk_picosecond_2017} plasmonics,~\cite{larrinzar_towards_2023,lanzillotti-kimura_polarization-controlled_2018,poblet_acoustic_2021,della_picca_tailored_2016}, polaritonics,~\cite{chafatinos_polariton-driven_2020,kuznetsov_electrically_2021,kobecki_giant_2022} and other related areas. To precisely tailor the phononic response and produce high-quality devices, expensive techniques such as molecular beam epitaxy and electron beam lithography are typically employed.~\cite{esmann_brillouin_2019-1,anguiano_micropillar_2017,poblet_acoustic_2021,della_picca_tailored_2016} However, there is a growing demand for cost-effective and reliable fabrication processes in engineering artificial materials for high-frequency acoustic applications. One promising solution is using nanoscale mesoporous materials, which possess a high surface-to-volume ratio and tailorable mesopores.~\cite{soler-illia_mesoporous_2006-1,gazoni_designed_2017-1, soler-illia_chapter_2021,hofmann_phonons_2017-1} These materials allow for chemical functionalization,~\cite{pizarro_droplets_2022,fuertes_photonic_2007-1} have versatile optical  applications,~\cite{pan_block_2011-1,auguie_tamm_2014-1} and offer a phononic response within the GHz range.~\cite{Benetti2018,Abdala2020,cardozo_de_oliveira_probing_2023,mechri_confined_2012,hernando_abad_sub-thz_2023}

In recent works, we reported the successful development of mesoporous silicon dioxide (SiO$_{2}$) and titanium dioxide (TiO$_{2}$) thin layers capable of supporting coherent acoustic modes spanning from 5 to 100 GHz, with Q-factors ranging from 5 to 17.~\cite{Abdala2020,cardozo_de_oliveira_probing_2023} Building upon this progress, we theoretically investigate a platform for engineering complex mesoporous systems specifically designed to support targeted high-frequency acoustic phonon modes, thus optimizing their performance for sensing applications. Our proposed structures consist of a mesoporous surface cavity placed atop an acoustic distributed Bragg reflector (DBR), enabling the confinement of phonons at 100 GHz. By incorporating environmental molecules into the pores and manipulating the elastic properties of the metamaterial, the mesoporous layer could potentially serve as the active element for sensing applications.~\cite{Benetti2018}

Our simulation results demonstrate the presence of robust coherent phonon signals at the desired frequency, offering the potential for achieving high Q-factors. This significant advancement paves the way for a promising platform in nanoacoustic sensing and reconfigurable optoacoustic nanodevices, all made possible through the utilization of soft and cost-effective fabrication methods.

\section{Methodology}

In a pump-probe coherent phonon generation experiment, standard technique in nanophononics, a strong laser pulse, namely the pump, interacts with the structure, triggering the phononic dynamics in the system. The modulation of the material refractive index due to these vibrations, i.e., the photoelastic interaction, takes place. Finally, a delayed probe reaching the structure measures transient reflectivity changes due to the presence of coherent acoustic phonons.~\cite{thomsen_coherent_1984,ruello_physical_2015} In this work, we simulate the generation and detection spectra of coherent acoustic phonons in such an experiment and test different sample designs. This is done by implementing the photoelastic model using the standard transfer matrix method for acoustic and optical fields.~\cite{lanzillotti-kimura_theory_2011,Fainstein2007} The theoretical approach employed here allows us to investigate multilayered structures with contrasting acoustic and optical impedances and test different parametrizations.~\cite{Fainstein2007} The material parameters used in the simulations are presented in Table~\ref{table_1}. 

\begin{table}[]
\caption{Optical and elastic properties of the studied materials for the numerical simulation.}
\label{table_1}
\resizebox{\columnwidth}{!}{%
\begin{tabular}{llllll}
\hline \rule{0pt}{2.6ex}
Material~~~~                                                    & Index of refraction~~           &  & Speed of sound (m/s)~~ &  & Density (g/cm$^{3}$) \\ \hline
\rule{0pt}{2.6ex}
dSiO$_{2}$                                                       & 1.538        &  & 5750    &  & 2.2         \\ \rule{0pt}{2.6ex}
dTiO$_{2}$                                                       & 2.56         &  & 6700    &  & 2.9         \\ \rule{0pt}{2.6ex}
\begin{tabular}[c]{@{}l@{}}mSiO$_{2}^*$ \\ (\textit{p} = 45\%)\end{tabular} & 1.323        &  & $\sim$  &  & $\sim$      \\ \rule{0pt}{2.6ex}
Nickel                                                      & 2.218+4.893$i$ &  & 5580    &  & 8.908       \\ \rule{0pt}{2.6ex}
Air                                                         & 1.0003      &  & 343     &  & 1.28x$10^{-3}$    \\ \rule{0pt}{2.6ex}
Water                                                       & 1.33         &  & 1480    &  & 0.997       \\ \rule{0pt}{2.6ex}
Glass                                                       & 1.538        &  & 5750    &  & 2.2         \\ \hline
\end{tabular}
}
* Mesoporous SiO$_{2}$. Refer to section~\ref{sct:performance} for details on the speed of sound and density.
\end{table}

We assume a white phonon spectrum propagating from the substrate towards the surface and compute the normalized stationary solutions for each phonon frequency. This allows us to calculate the acoustic phonon transmission spectrum and determine the frequency-dependent acoustic field distribution along the structure. Analogously, the incident laser electric field profile can be calculated by considering Maxwell's equations for electromagnetic waves.~\cite{matsuda_reflection_2002} We assume standard boundary conditions, and a monochromatic laser to simplify the formulation. Finally, we determine the acoustic phonon generation spectrum ($G(\omega)$) with an overlap integral of the incident pump's electric field ($E(\nu,z)$), the strain field ($\eta(\omega,z)$), i.e., the derivative of the acoustic displacement, and the material-dependent transduction constant ($k(z)$), as follows: 
\begin{equation}
G(\omega) = \int k(z)\eta(\omega,z)|E(\nu,z)|^2dz,
\label{eq1}
\end{equation}
where $\omega$ and $\nu$ correspond to the acoustic and optical frequencies, respectively. The coherent phonon detection spectrum, which can be related to the modulation of the probe reflectivity ($\Delta R(\omega)$), is then calculated by considering the photoelastic interaction in the material. This is done by calculating an overlap integral similar to equation~\ref{eq1}, but taking into account the generation spectrum and the probe's complex electric field, in the form:
\begin{equation}
\Delta R(\omega) = \int G(\omega) p(z)\eta(\omega,z)E(\nu,z)^2dz
\label{eq2}
\end{equation}
Additionally, we can determine the surface displacement by multiplying the phonon generation spectrum with the transmission spectrum:
\begin{equation}
d(\omega) = G(\omega) T(\omega)
\label{eq3}
\end{equation}
Experimentally, the surface displacement can be measured by implementing an interferometer in the pump-probe setup, which involves sensitive alignment procedures.~\cite{perrin_interferometric_1999} While it is easier to probe changes in reflectivity, the analysis of surface displacement becomes necessary depending on the specific sample under investigation.~\cite{Abdala2020,cardozo_de_oliveira_probing_2023}

It is worth noting that our analysis assumed that the optical absorption and coherent phonon generation are limited to the nickel transducer layer.

\section{Design of mesoporous surface acoustic resonators}

Mesoporous materials are known for their ability to adsorb chemical compounds into the mesopores.~\cite{ruminski_humidity-compensating_2008, sansierra_detection_2019} This leads to effective functionalization of the materials, that can be sensitive to different external stimuli. In the context of nanoacoustics, the chemical infiltration modifies the overall elastic properties of the material, such as mass density ($\rho$) and speed of sound ($v$). The simplest acoustic resonator can be engineered by embedding a mesoporous thin film (MTF) between two contrasting acoustic impedance ($Z$) layers. Depending on the relative impedances of the layers, the resonator will support frequency modes according to either
\begin{subequations}
\begin{equation}
f_{n} = \dfrac{nv}{2d}
\label{eq:ac_resonance}
\end{equation}
or
\begin{equation}
f_{n} = \dfrac{(2n-1)v}{4d},
\end{equation}
\end{subequations}
where $d$ and $n=1,2,3...$ are associated with the mesoporous thickness and mode order, respectively.  Here, the acoustic resonances in the mesoporous materials can be determined with the eq.~\ref{eq:ac_resonance} The acoustic impedance mismatch between the layers, given by the contrast in mass density and speed of sound of the involved materials, can account for enhanced transmission or reflection of the sound waves at the interfaces. Therefore, chemical infiltration into the mesopores can significantly change the resonance frequency -by changing the $v$, and the Q-factor of the confined modes -by changing $v$ and $\rho$. A simple approach could be to submit the device to different relative humidity (RH) conditions, which could be extended for sensing different gases.

In fact, phonon confinement in mesoporous thin films based on SiO$_{2}$ and TiO$_{2}$ has been recently reported in pump-probe experiments.~\cite{Abdala2020,cardozo_de_oliveira_probing_2023} In such works the mesoporous layer was embedded between a glass substrate and a nickel thin-film that acts as the acousto-optical transducer. Two major constraints arise from this design: the mesoporous thin film is covered by other layers, which prevents efficient adsorption into the mesopores; and the leakage of the acoustic waves towards the substrate reduces the Q-factor of the resonator. 

\begin{figure}[h!]
	\par
	\begin{center}
        \includegraphics[scale=0.34]{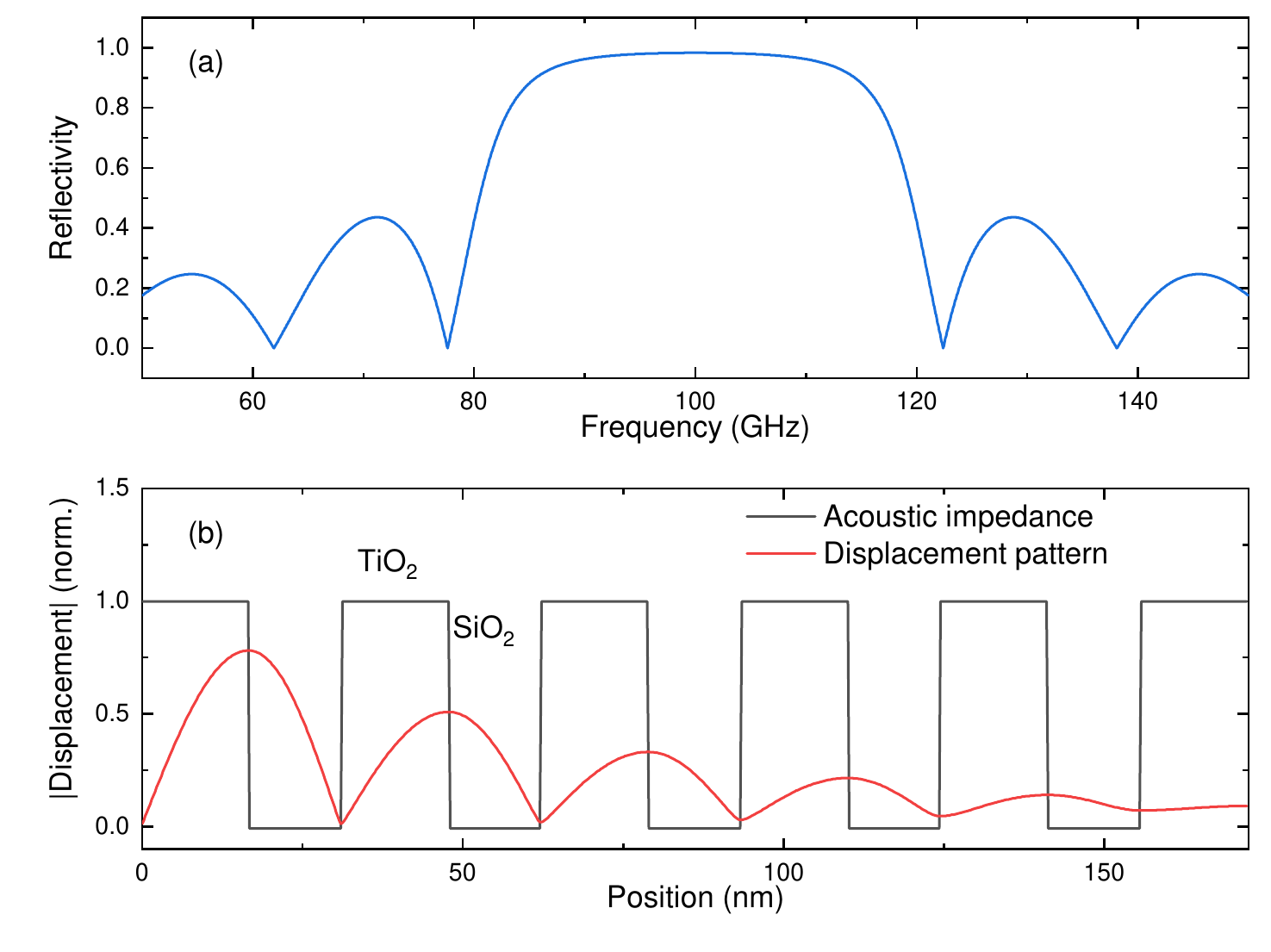} 
	\end{center}
	\par
	\vspace{-0.5cm} 
    \caption{(a) Response of the acoustic DBR formed by dense SiO$_{2}$/TiO$_{2}$ layers, showing a stop band centered at 100 GHz. (b) The acoustic displacement pattern at 100 GHz (red) and the acoustic impedance (black) along the structure.}
	\label{schematics}
\end{figure}

To overcome these limitations we propose a device design composed of the MTF as the outermost layer in the structure, followed by the nickel transducer and an acoustic distributed Bragg reflector (DBR) based on dense SiO$_{2}$/TiO$_{2}$ on top of a glass substrate. Since the mesoporous SiO$_{2}$ is a transparent material for the studied optical range (near infrared), the phonon generation takes place in the nickel layer upon light absorption, mainly via photoinduced thermoelasticity.~\cite{ruello_physical_2015} The acoustic DBR designed at a frequency $f$ has an acoustic stop-band that reflects the frequency components within this band, as shown in Figure~\ref{schematics}, for $f=100$~GHz. On the other hand, GHz acoustic phonons are entirely reflected at the mesoporous/air interface, as they do not propagate into the air. With that, by properly engineering the thicknesses of the layers, it is possible to tune the DBR stop-band and develop a surface acoustic cavity confining phonons at the mesoporous layer.

\begin{figure}[h!]
	\par
	\begin{center}
        \includegraphics[scale=0.28]{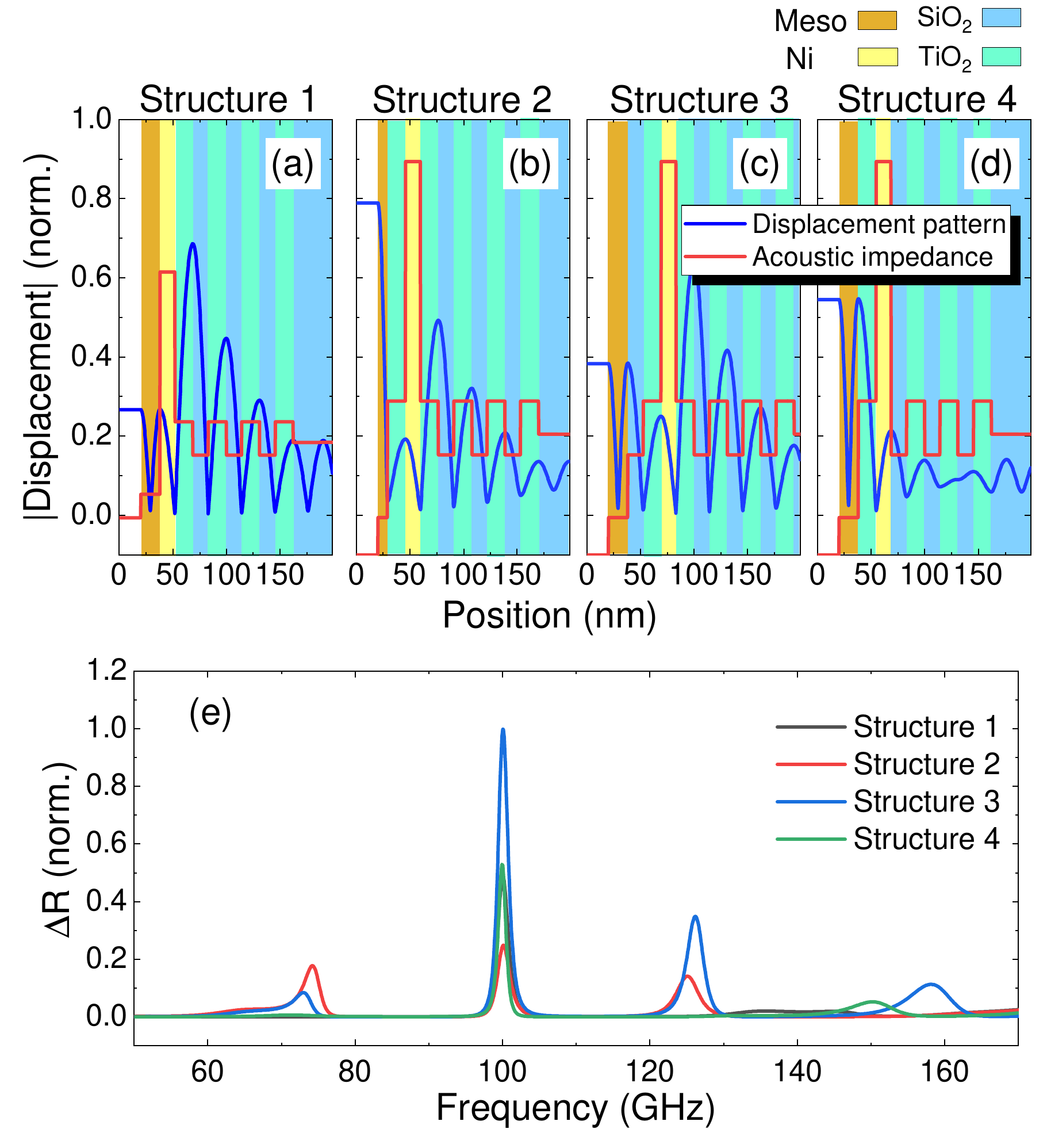} 
	\end{center}
	\par
	\vspace{-0.5cm} \caption{ (a-d) Normalized displacement pattern (blue) and acoustic impedance (red) of the surface cavity structures 1-4. The layers of mesoporous, nickel, SiO$_{2}$ and TiO$_{2}$ are represented by the colors orange, yellow, blue, and green, respectively. (e) Reflectivity spectra for the structures presented in panels (a-d). The inset represents an enlargement of the peak at 100 GHz.}
	\label{structures}
\end{figure}

The acoustic resonators (ARs) proposed here are formed by four different elements. We engineer and numerically investigate four distinct variations of ARs, shown in Figs~\ref{structures}(a-d). These variations involve different combinations of layers. Our objective is to identify the optimal configuration that exhibits the strongest correlation between acoustic resonance and the infiltration of liquids into the mesopores. The devices are formed by a glass substrate, followed by a 3-periods-DBR of TiO$_{2}$/SiO$_{2}$ ($\lambda/4$,$\lambda/4$) designed to operate at 100 GHz, where $\lambda$ corresponds to the acoustic wavelength at the design frequency. The structures are named AR1, 2, 3 and 4. ARs 1-3 have an extra $\lambda/4$ TiO$_2$ layer before the $\lambda/4$ Ni transducer. After the nickel, the four structures have a combination of zero, one, or two $\lambda/4$ dense layers before finalizing with the mesoporous layer. With these configurations, the mesoporous thickness leading to phonon confinement at 100 GHz is $\lambda/2$ for AR1, AR3, and AR4, and $\lambda/4$ for AR2. 

Figure~\ref{structures}(e) displays the simulated phononic spectra of all the structures, considering the modulation of the refractive index of the Nickel layer, with a pronounced acoustic mode at 100 GHz. This result supports the use of a reflectometric experimental setup for detecting the acoustic resonances in the mesoporous layer. The acoustic resonance detection performed by the Ni layer indicates that its proximity to the mesoporous layer is not a critical requirement. The peaks at 75 GHz and 125 GHz indicate a stop-band spanning over $\sim$50 GHz. The acoustic resonance in AR2 has the weakest response. To investigate the region in which the phonons are confined in the structures, the displacement pattern at 100 GHz is plotted, overlapped with the device profiles (blue curves of Figs.~\ref{structures}(a-d)), the maximum displacement indicates the region of phonon confinement. In such analysis, AR1 reveals as the only device in which the phonon confinement is not at the mesoporous layer; however, it has the strongest displacement among all. AR3 presents maximum amplitude in two different regions of the structure, which indicates the formation of a surface and a cavity mode. Finally, AR2 and AR4 have a maximum displacement at the mesoporous layer.

\begin{figure}[h!]
	\par
	\begin{center}
        \includegraphics[scale=0.33]{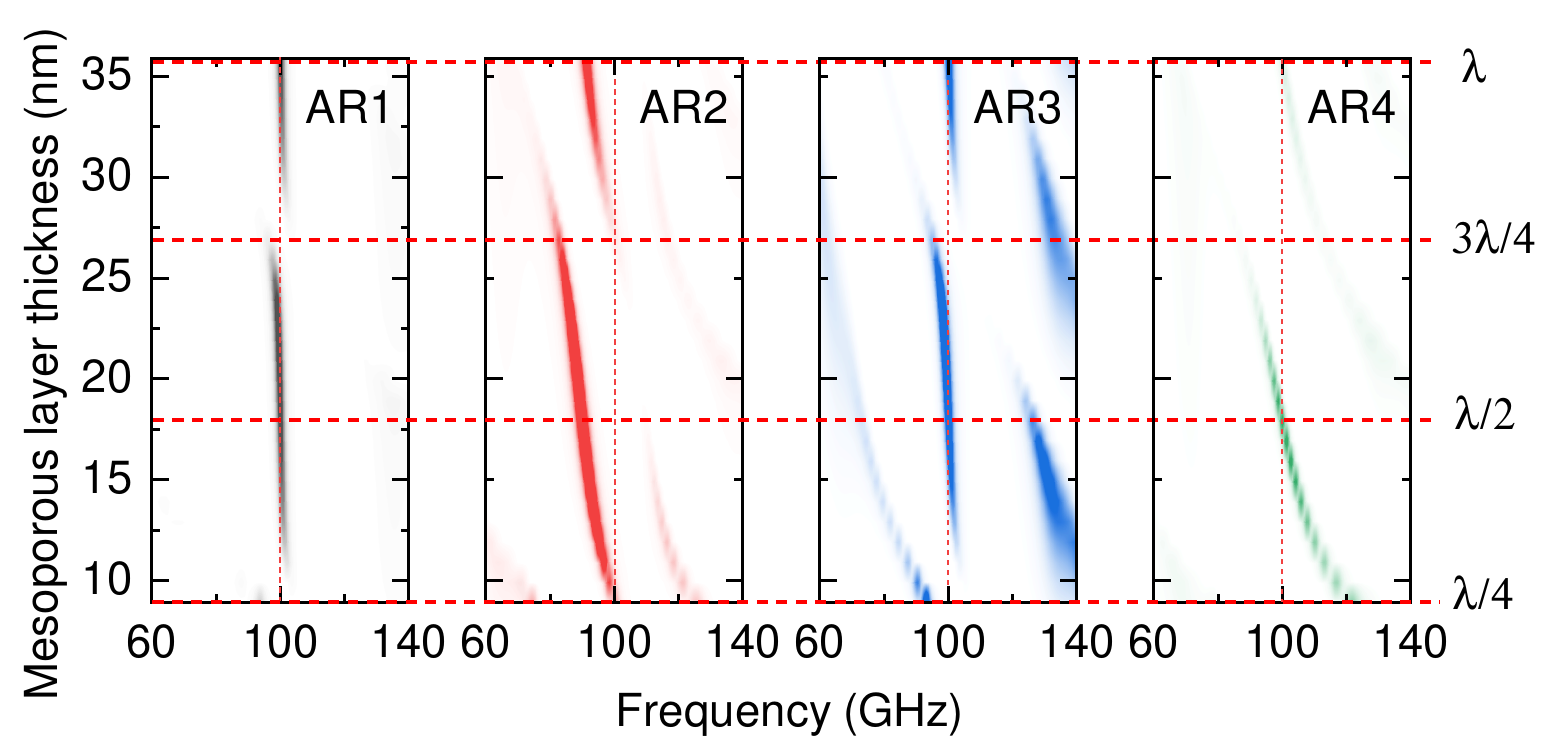} 
	\end{center}
	\par
	\vspace{-0.5cm} \caption{ Photoelastic interaction as a function of the mesoporous layer thickness and frequency for AR1, AR2, AR3, and AR4. Horizontal dashed lines represent the key layer thicknesses in steps of a quarter of the phonon wavelength. Vertical dashed lines are guides to the design frequency of 100 GHz.}
	\label{colorplot}
\end{figure}

Usually, the fabrication of multilayered structures is susceptible to thickness fluctuations. When engineering acoustic resonators, the precise definition of the layer thicknesses is essential, and the thickness variation might compromise the phonon confinement. Notwithstanding, we study the photoelastic interaction as a function of the mesoporous layer thickness, in terms of phonon wavelength $\lambda$, ranging from $\lambda/4$ to $\lambda$, and the results are presented in Figure~\ref{colorplot}. 

The mesoporous thicknesses for which the resonators exhibit the mode frequency at 100 GHz are $\lambda/2$ for AR1, AR3 and AR4, and $\lambda/4$ for AR2, in agreement with the designed structures. Furthermore, we can identify how sensitive the acoustic resonances are with the mesoporous thickness. By changing the layer thickness from $\lambda/4$ to $3\lambda/4$, the acoustic resonances of AR2 and AR4 span over $\sim$~20 and $\sim$~30 GHz, respectively. Conversely, AR1 and AR3 have minimum dependence on the mesoporous thickness, as the phonon confinement region is located between the Ni and the DBR. It is worth noting that the modes within the stop band persist, even if the resonance deviates from the specified 100 GHz.

\section{Optimization of the devices}

\begin{figure}[h!]
	\par
	\begin{center}
        \includegraphics[scale=0.28]{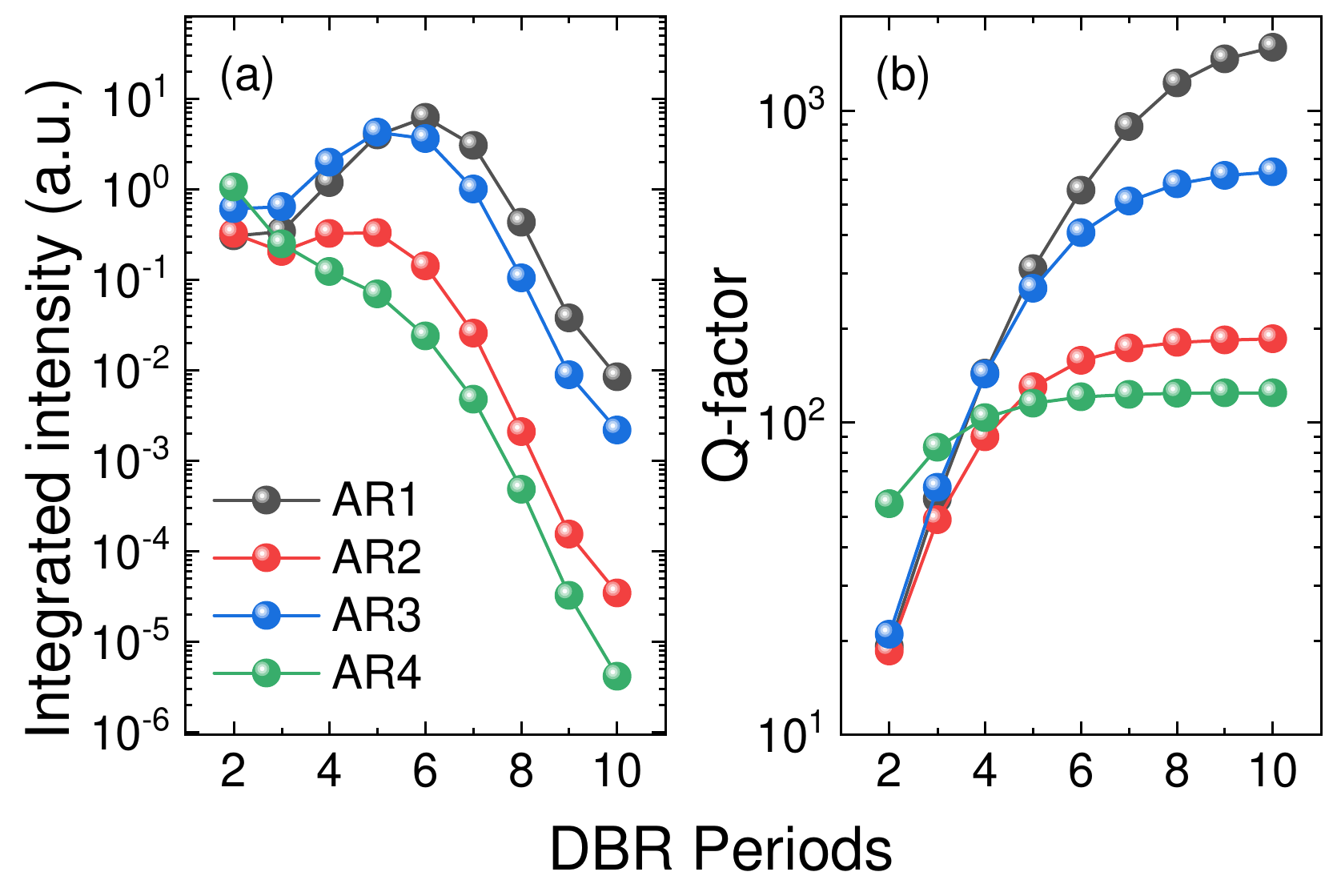} 
	\end{center}
	\par
	\vspace{-0.5cm} \caption{(a) Integrated intensity and (b) quality factor of the acoustic mode at 100 GHz as a function of the number of DBR periods for the four proposed resonators.}
	\label{DBR_periods}
\end{figure}

To determine the optimal responsivity of the proposed devices, we simulated the photoelastic interaction of structures with a different number of the dense SiO$_{2}$/TiO$_{2}$ bilayers composing the DBR, ranging from 2 to 10 periods, for the 4 structures. Afterward, the mode at $f=100$~GHz in the full pump-probe simulation is fitted with a Lorentzian function, and both the integrated intensity and full width at half maximum (FWHM) are extracted. The results are shown in Figure~\ref{DBR_periods}.

The integrated intensity (Fig.~\ref{DBR_periods}(a)) is strongly dependent on the number of DBR periods, for all the ARs, changing up to five orders of magnitude over the calculated range. Maximum intensity is observed at 6 periods for AR1 and 5 periods for AR2 and AR3, and AR4 displays the maximum value at 2 periods. The quality factor (Q-factor), shown in Fig.~\ref{DBR_periods}(b), is calculated by the ratio between the mode frequency (100 GHz) and the extracted FWHM. All the resonators present an increase of the Q-factor as the number of DBR periods increases, as expected, showing a tendency for stabilization at higher periods. At 10 periods, the highest Q-factor, of $\sim$1000, is for AR1, followed by AR3 ($\sim$550), AR2 ($\sim$180) and AR1 ($\sim$150).

\section{Performance of the devices}\label{sct:performance}

\begin{figure}[h!]
	\par
	\begin{center}
        \includegraphics[scale=0.28]{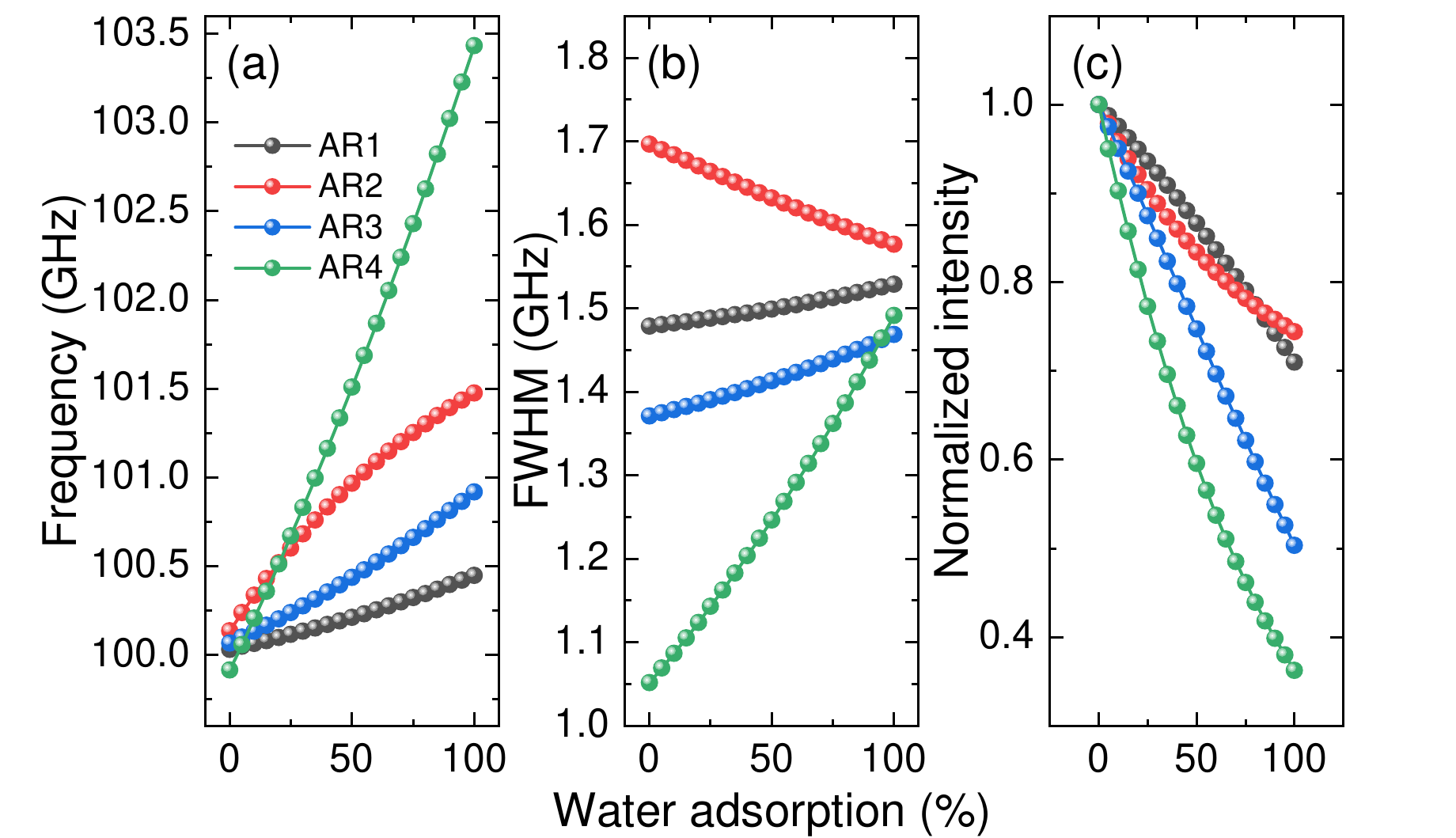} 
	\end{center}
	\par
	\vspace{-0.5cm} \caption{(a) Frequency shift, (b) FWHM, and (c) normalized intensity of the mode at 100 GHz as a function of water adsorption for the four proposed structures.}
	\label{sensor}
\end{figure}

Finally, we investigate the performance of the acoustic resonators over liquid infiltration into the nanopores. It is still unclear how acoustic resonators in the GHz range respond to the density and speed of sound changes in the MTF under liquid infiltration.~\cite{Abdala2020,cardozo_de_oliveira_probing_2023} High-frequency coherent acoustic phonons are unable to penetrate into liquids or gases; therefore, they remain entirely on the solid matrix without penetrating into the nanopores, and no response would be perceived. However, recent studies indicate that in nanoscale mesoporous materials, liquids, such as water, condensate at the boundaries of the matrix and behave as solid elements.~\cite{velasco_2017, gonzalez_solveyra_2013}. This allows high-frequency phonon penetration, leading to a change in the effective density, as well as in the speed of sound. The pore-load modulus, a quantity associated with the pressure inside the pores under liquid infiltration, has also been studied in silicon mesoporous membranes as a function of the material porosity.~\cite{gor_elastic_2015} Elucidating such effects is out of the scope of this work. Hereupon, in order to verify the effect of density and speed of sound change in the material, we consider weighted averages between the solid matrix, porosity, and humidity, while keeping the other parameters (such as mesoporous layer thickness) constant, according to the equation
\begin{equation}
X = X_{\textrm{solid}}\left(1-p\right) + \left[X_{\textrm{air}}\left(1-h\right)+X_{\textrm{water}}h\right]p,
\label{weight_average}
\end{equation}
where $X$ represents either the density or the speed of sound, whereas $p$ and $h$ correspond to the porosity and humidity ratios. We then perform the simulation for all the acoustic resonators with a constant porosity of 40~$\%$, and variable humidity between 0 and 100~$\%$, in which the reference 100 GHz mode is calculated at 0 $\%$ humidity. By fitting a Lorentzian function on the acoustic resonance, we extract and analyze the frequency position, FHWM, and normalized intensity variation. The results for each of the peak properties are respectively depicted in Figs.~\ref{sensor}(a), (b), and (c).

The water adsorption on the mesoporous layer influences the three peak parameters. All the resonators undergo a blueshift in frequency at higher humidity (Figs.~\ref{sensor}(a)), which is related to the increase in the speed of sound. As the sound speed in the water is higher, the weighted average increases and the resonator confines higher frequencies at the same region. It is worth noticing that AR1 is the least sensitive, with a frequency shift $\Delta f \sim 0.5$ GHz from 0 to 100 $\%$ RH, whereas AR4 is the most sensitive ($\Delta f \sim 3.5$ GHz). This can be justified by the fact that AR1 is the only resonator in which the confined mode is not at the mesoporous layer, whereas AR4 displays clear confinement at this layer, as seen in Figs.~\ref{structures}(a) and (d). The highest sensitivity for AR4 and lowest for AR1 is also true for the analysis of FWHM and integrated intensity. The linewidth of the resonance broadens for the resonators AR1, AR3, and AR4, whereas it narrows for AR2, as shown in Fig.~\ref{sensor}(b). Liquid infiltration into the mesopores also leads to a decrease in the integrated intensity of the peak at 100 GHz for all the resonators (Fig.~\ref{sensor}(c)), which reach values of $\sim 0.68$, $\sim 0.77$, $\sim 0.45$, and $\sim 0.40$ compared to the maximum intensity at 0 $\%$ RH, for each of the resonators AR1, AR2, AR3, AR4, respectively. This reduction in intensity comes from the fact that the mesoporous acoustic impedance increases under liquid infiltration, leading to a less efficient coupling of the acoustic waves to the transducer.

The three peak parameters of the four resonators show a nearly linear dependence on water adsorption, which is an essential aspect of the calibration and scale of sensors. The design of AR4 is considered to provide the best sensitivity. Despite the observed differences, the four structures proposed in here are potential candidates for sensing applications. Further experimental efforts should explore this potential in detail. Additionally, the scope of the study could be expanded to include other systems, such as the detection of different gases with distinct densities, which would likely yield varied responses.

\section{Conclusion}

We have introduced a new design for optoacoustic sensing devices featuring mesoporous thin film as the active component. By infiltrating the layer with liquid, the elastic properties of the thin film can be gradually modified. As a result, an acoustic resonator employing these materials experiences an overall alteration of its resonance characteristics. We theoretically demonstrated the promising potential of these devices for environmental sensing. They can exhibit high-quality factors of up to 1000, a nearly linear response of peak parameters to changes in humidity, and a significant peak intensity variation of up to 60~$\%$. The small size of the resonator and the short wavelength of GHz phonons offer a notable advantage over other sensors, as they enable significantly faster responsiveness. This means that a smaller volume of liquid infiltration is needed to achieve effective sensing. However, the requirement for all-optical experiments to investigate high-frequency acoustic phonons remains a significant roadblock for practical applications. This challenge could potentially be overcome by combining electrically contacted bulk acoustic wave resonators for efficient transduction.~\cite{crespo-poveda_ghz_2022,machado_generation_2019} Nevertheless, our findings shed light on a novel framework for nanoacoustic sensing and adaptable devices, utilizing cost-effective fabrication techniques.

\section{Acknowledgment}
The authors acknowledge funding by the ECOS-Sud Program through the project PA19N03 (TUNA-Phon). NDL-K acknowledges funding by the European Research Council Consolidator Grant No. 101045089 (T-Recs). GJAASI acknowledges ANPCyT for projects PICT 2017-4651, PICT-2018-04236, and PICT 2020-03130. PV acknowledges CONICET for the Doctoral Fellowship.

\bibliography{design_mesoporous}

\end{document}